\documentclass[amsmath,amssymb, aps, pra,reprint,footinbib,floatfix,superscriptaddress]{revtex4-2}

\usepackage{float}
\usepackage{amsmath,amssymb}
\usepackage[english]{babel}
\usepackage{upgreek}
\usepackage{dsfont}
\usepackage{graphicx}
\usepackage{color}
\usepackage{bbold}
\usepackage{hyperref}
\usepackage[authormarkup=none]{changes}

\begin{document}
\title{Passive quantum phase gate for photons based on three level emitters}

\author{Bj\"{o}rn Schrinski}
\affiliation{Center for Hybrid Quantum Networks (Hy-Q), The Niels Bohr Institute, University of Copenhagen, Blegdamsvej 17, 2100 Copenhagen, Denmark}
\author{Miren Lamaison}
\affiliation{Center for Hybrid Quantum Networks (Hy-Q), The Niels Bohr Institute, University of Copenhagen, Blegdamsvej 17, 2100 Copenhagen, Denmark}
\author{Anders S. S\o rensen}
\affiliation{Center for Hybrid Quantum Networks (Hy-Q), The Niels Bohr Institute, University of Copenhagen, Blegdamsvej 17, 2100 Copenhagen, Denmark}

\begin{abstract}

We present a fully passive method for implementing a quantum phase gate between two photons travelling in a one-dimensional wave guide. The gate is based on  chirally coupled emitters in a three-level $V$ configuration, which only interact through the photon field without any external control fields. We describe the (non-)linear scattering of the emerging polariton states and show that for near resonant photons the scattering dynamics directly implements a perfect control phase gate between the incoming photons in the limit of many emitters. For a finite number of emitters we show that the dominant error mechanism can be suppressed by a simple frequency filter at the cost of a minor reduction in the success probability. We verify the results via comparison with exact scattering matrix theory and show that the fidelity can reach values $\mathcal{F}\sim99\%$ with a gate success probability of $>99\%$ for as few as 8 emitters.   
\end{abstract}

\maketitle

\emph{Introduction--} 
Light has  great advantages as carrier of quantum information since it travels very fast. Furthermore, light is largely invulnerable to decoherence even at room temperature since  photons are rather absorbed than decohered. On the other hand, photons practically do not interact with each other  \cite{d2013observing}, which makes it highly challenging to implement  quantum gates, the key building blocks  for quantum information processing. Alternative strategies have been introduced by combining sources of single or entangled photons with measurement and feedback \cite{knill2001scheme,briegel2009measurement,gimeno2015three,azuma2015all}. These techniques are, however, inherently probabilistic and come with a large overhead in resources \cite{browne2005resource,pant2017rate,pant2019percolation}. 
One way to avoid these complications is to achieve an indirect interaction
via coupling to non-linear matter.
There have been several ideas for how to achieve gate operations in such systems: 
Strong coupling of photons to individual systems, such as  atoms in optical cavities \cite{chudzicki2013deterministic} or optomechanical resonators \cite{gea2014conditional}  can induce phase gates  with relatively high fidelity   for small phase shifts but not for large shifts due to pulse distortions. These distortions can be avoided by  mapping the photons into excitations of optical cavities \cite{heuck2020controlled} or individual atoms, either coupled to wave guides \cite{chang2007single} or optical cavities   \cite{duan2004scalable,duan2005robust}, and subsequently scattering a second photon of the system. Similarly, photons can be stored in ensembles of  atoms either in free space \cite{mavsalas2004scattering,friedler2005long,andre2005nonlinear,gorshkov2010photonic,gorshkov2011photon,thompson2017symmetry,iakoupov2018controlled} or optical cavities \cite{das2016photonic}. Gates can then be implemented by relying on direct interaction of the stored excitations in optical lattices \cite{mavsalas2004scattering,gorshkov2010photonic} or through scattering of a second photon pulse and exploiting  Rydberg interactions \cite{friedler2005long,gorshkov2011photon,das2016photonic} or stationary light effects \cite{andre2005nonlinear,iakoupov2018controlled}. Common to these proposals, however, is that they rely on precisely timed lasers pulses for photons storage, which complicates their implementation. 

\begin{figure}[b]
  \centering
  \includegraphics[width=0.45\textwidth]{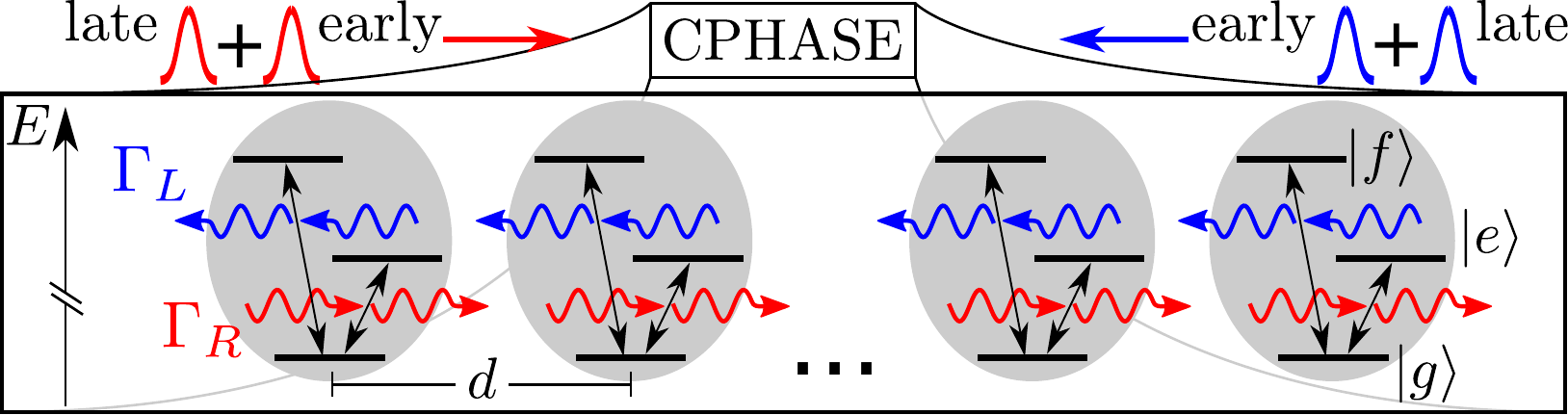}
  \caption{Photonic phase gate for counterpropagating wave packets in a one dimensional waveguide, coupled to three level dipole emitters with distance $d$ to the next neighbours. The transition $|g\rangle\to|e\rangle$ ($|g\rangle\to|f\rangle$) in the three-level systems couple chirally to the right (left) traveling waveguide modes with  coupling rate $\Gamma_R$ ($\Gamma_L$). }
\label{fig:Waveguide}
\end{figure}

Photon gates not depending on precise timing of pulses have been  developed using counter propagating pulses in Rydberg EIT systems \cite{he2014two} or through static cross-Kerr interactions between emitters coupled to different one-dimensional waveguides \cite{brod2016two,brod2016passive}. Inspired by the latter, we propose an experimentally viable setup for implementing quantum gates based on counterpropagating photons  in  waveguides chirally coupled to quantum emitters, such that the emitters decay by only emitting light in one direction \cite{lodahl2017chiral,hauff2022chiral}.  Such couplings have previously been demonstrated \cite{luxmoore2013interfacing,sollner2015deterministic,coles2016chirality,guimond2016chiral,barik2020chiral} and also used for (non-)linear operations   \cite{volz2014nonlinear,shomroni2014all}.
Here we exploit the chiral coupling to achieve a simple implementation of a cascaded quantum system \cite
{stannigel2012driven,ramos2014quantum,pichler2015quantum} without the need for multiple Kerr-circulators or multiple arrays of chirally coupled emitters as in Ref.~\cite{brod2016passive}.  We furthermore avoid any complication of engineering interactions between emitters by basing our setup on three level dipole emitters in the $V$ configuration coupled to a one-dimensional waveguide,  see Fig.~\ref{fig:Waveguide}. This configuration contains an inherent Kerr non-linearity since the emitter cannot be excited twice, and the only required coupling is between the emitters and the photons.
In contrast to previous  proposals, 
the assumed setup thus enables the implementation of a gate completely passively without the need for any control fields, Rydberg EIT or other interactions while still promising excellent fidelity.

We will show that this setup implements any desired two-photon phase gate controlled by the energy of the photons. If the photons are on resonance the underlying  physics can be sketched quite easily: Photons on resonance receive a phaseshift of $\pi$ every time they scatter off an emitter \cite{chang2012cavity}. Because of the chiral coupling the photons cannot be reflected but rather travel from emitter to emitter. 
This leads to an overall phase of $(-1)^N$ accumulated by photons leaving on the opposite side of the array with $N$ the number of emitters. If two counter-propagating photons are  inside the emitter chain at the same time they have to pass one another at some point but cannot simultaneously excite the very emitter at which they  cross, leading to an additional phase factor of $(-1)$. In a time-binned superposition where the late state of the right propagating photon arrives simultaneously with the early state of the left propagating photon (see Fig.~\ref{fig:Waveguide}) the emitter array therefore implements the transformation 
\begin{align}
&|\mathrm{early},\mathrm{early}\rangle&&\to\quad|\mathrm{early},\mathrm{early}\rangle\nonumber\\
&|\mathrm{late},\mathrm{late}\rangle&&\to\quad|\mathrm{late},\mathrm{late}\rangle\nonumber\\
&|\mathrm{early},\mathrm{late}\rangle&&\to\quad|\mathrm{early},\mathrm{late}\rangle\nonumber\\
&|\mathrm{late},\mathrm{early}\rangle&&\to\quad-|\mathrm{late},\mathrm{early}\rangle,
\end{align}  
where the first (last) entry in the state refer to the right(left) propagating photon. 
This creates a controlled $Z$-phase gate flipping the sign in case of two photons being simultaneously in the system.

The description above relies on photons simultaneously being resonant and  inside the emitter chain. Fundamentally this cannot be achieved with a single emitter.  The finite size of the photon wave packets thus   leads to  degradation of the gates due to dispersion, a smeared out phase and inelastic scattering \cite{shapiro2006single,gea2010impossibility,fan2010input,xu2013analytic,chudzicki2013deterministic,gea2014conditional,xu2015input,caneva2015quantum,ke2019inelastic,schrinski2021polariton} which is the reason why similar proposals only achieve high fidelities for only small phases \cite{chudzicki2013deterministic,gea2014conditional}. Multiple emitters are thus required to ensure that  the photons meet each other inside the chain despite the large time uncertainty from being narrow in frequency.   In the following we give a concise analytical description of the process, which enables us to address this and other imperfections which potentially compromise the fidelity of the gate.  

\emph{System--}The system in question consists of a one-dimensional waveguide coupled to many three level emitters with next neightbour distance $d$, see Fig.~\ref{fig:Waveguide}.
We consider two counterpropagating photonic modes with field operators $\mathcal{E}_R(z)$ and $\mathcal{E}_L(z)$, respectively, with commutation relation $[\mathcal{E}_{R/L}(z_1),\mathcal{E}^\dagger_{R/L}(z_2)]=c\,\delta(z_1-z_2)$, where $c$ is the group velocity of light in the waveguide (assumed identical in both directions). Transitions between  different levels of an emitter are described by operators $\sigma_{ab}^{(\mu)}=|a\rangle_\mu\langle b|_\mu$. 
Applying the rotating wave approximation,  and rescaling the photon energy by the transition energies 
so that all energies are relative to resonance, we have the Hamiltonian $\mathsf{H}_\mathrm{tot}=\mathsf{H}_\mathrm{p}+\mathsf{H}_\mathrm{int}$ with 
\begin{align}
\mathsf{H}_\mathrm{p}=-i\hbar \int \mathrm{d}z\left[\mathcal{E}_R^\dagger(z)\partial_z\mathcal{E}_R(z)-\mathcal{E}_L^\dagger(z)\partial_z\mathcal{E}_L(z)\right]
\end{align}
and
\begin{align}
\mathsf{H}_\mathrm{int}=\sum_\mu\left[\mathcal{E}_R(z_\mu)\sigma^{(\mu)}_{eg} g_R+\mathcal{E}_L(z_\mu)\sigma^{(\mu)}_{fg} g_L\right] +\mathrm{h.c.}.
\end{align}

For our purposes we assume an identical coupling rate to the left and
right travelling modes $g_R=g_L=g_0$. Further, we assume the photons to travel quasi instantaneously from emitter to emitter relative to the lifetime of excitations. This allows us to eliminate the photonic degrees of freedom \cite{caneva2015quantum} and to describe the dynamics within the emitter array on the level of polaritons, i.e.\,coupled light-matter states.
The effective Hamiltonian  acting on the polaritons then reads
\begin{align}\label{eq:effHam}
\mathsf{H}=&-i\frac{\Gamma_0}{2}\sum_\mu (\sigma^{(\mu)}_{eg}\sigma^{(\mu)}_{ge}+\sigma^{(\mu)}_{fg}\sigma^{(\mu)}_{gf})\nonumber\\
&-i\Gamma_0\sum_{\mu<\nu}e^{ik_0|z_\mu-z_\nu|}(\sigma^{(\nu)}_{eg}\sigma^{(\mu)}_{ge}+\sigma^{(\mu)}_{fg}\sigma^{(\nu)}_{gf}) ,
\end{align}
with $k_0$ the wavenumber at the resonance frequency and $\Gamma_0=2\pi g^2_0$ the decay rate of the emitter into the waveguide. This effective Hamiltonian is non-hermitian, reflecting that excitations can leave the system at the edges. 

We are interested in a setup including many emitters and thus solve \eqref{eq:effHam} in the limit of infinitely many interaction sites, where we can neglect losses at the edges. 
For a single excitation propagating to the right this leads to a dispersion relation
\begin{align}\label{eq:DispersionRelation}
\omega_1=\omega(k_1)=-\frac{\Gamma_0}{2}\frac{\cos[(k_1-k_0)d/2]}{\sin[(k_1-k_0)d/2]}
\end{align}
for the momentum state excited by $\sigma^{k_1}_{eg}=\sqrt{d}\sum_\mu e^{ik_1z_\mu}\sigma^\mu_{eg}$ (and analogously $\omega_2=\omega(-k_2)$  for the left travelling one).  This results in an individual phase $\varphi_{\ell}=kNd$ accumulated while passing the whole array of emitters, with 
\begin{align}\label{eq:IndividualPhase}
e^{i\varphi_{\ell}}=\left(\frac{i\omega_\ell+\Gamma_0/2}{i\omega_\ell-\Gamma_0/2}\right)^N,
\end{align}
where $\ell\in 1,2$. This coincides with the exact result \cite{fan2010input} for finite chains of $N$ emitters and gives a phase shift of $(-1)^N$ when on resonance ($\omega_\ell=0$).

In the case of two polaritons 
 we can solve the scattering problem by switching to center-of-mass momentum $K=(k_1+k_2)/2$ and  relative momentum $q=(k_1-k_2)/2$ 
as well as the center-of-mass $z=(z_1+z_2)/2$ and relative position $\Delta=z_2-z_1$ of the two excitations \cite{Bakkensen2021photonic,schrinski2021polariton,supp}. The eigenstates of the Hamiltonian \eqref{eq:effHam} with eigenenergies $\omega_{q,K}=\omega_1+\omega_2$ then read
\begin{align}\label{eq:effHamEigenstate}
|\psi_{q,K}\rangle=\sum_{z,\Delta}f(q,\Delta)e^{2iKz}\sigma^{((z-\Delta/2)/d)}_{eg}\sigma^{((z+\Delta/2)/d)}_{fg}|0\rangle,
\end{align}
with
\begin{align}
f(q,\Delta)=
\begin{cases}
e^{iq\Delta}\quad&\mathrm{for}\quad\Delta>0\\
t_\mathrm{el}e^{iq\Delta}+t_\mathrm{in}e^{iq'\Delta}\quad&\mathrm{for}\quad\Delta<0,
\end{cases}
\end{align}
see Ref.  \cite{supp} for details of this and  the calculations below.

This expression describes elastic scattering with amplitude 
\begin{align}\label{eq:TransElastic}
t_\mathrm{el}=|t_\mathrm{el}|e^{i\varphi_{1,2}}=1 -\frac{2\Gamma_0(\Gamma_0-i\omega_1-i\omega_2)}{\Gamma_0^2+2\omega_1^2+2\omega_2^2}
\end{align}
preserving the relative momentum $q$ due to the chirality of the process. In addition, we  have an amplitude 
\begin{align}\label{eq:TransInelastic}
t_\mathrm{in}=\frac{2i\Gamma_0(\omega_1-\omega_2)^2}{(\omega_1+\omega_2-\Gamma_0)(\Gamma_0^2+2\omega_1^2+2\omega_2^2)}
\end{align}
for scattering into a degenerate $q'$ with $\omega_{q,K}=\omega_{q',K}$. This redistributes the energies of the outgoing photons 
\begin{align}\label{eq:EnergiesInelasticScattering}
\omega_1'=&\frac{\Gamma_0^2+2(\omega_1+\omega_2)\omega_2}{2(\omega_2-\omega_1)}\nonumber\\
\omega_2'=&-\frac{\Gamma_0^2+2(\omega_1+\omega_2)\omega_1}{2(\omega_2-\omega_1)}.
\end{align} 
while preserving the total energy $\omega_1+\omega_2=\omega_1'+\omega_2'$. We hence refer to it as  inelastic scattering. Importantly, these fulfill the continuity equation $|t_\mathrm{el}|^2+|t_\mathrm{in}|^2v_{q',K}/v_{q,K}=1$ where $v_{q,K}=\partial_q \omega_{q,K}$. 

From the scattering amplitudes (\ref{eq:TransElastic},\ref{eq:TransInelastic}) we observe that with both incoming photons on resonance ($\omega_1=\omega_2=0$) we get $t_\mathrm{el}=-1$, $t_\mathrm{in}=0$ corresponding to a perfect $Z$-gate as described in the introduction. Further, by going off-resonance while keeping $\omega_1=\omega_2=\omega$ we can ensure elastic scattering while achieving any desired phase shift 
\begin{align}\label{eq:ScatteringPhaseshift}
t_\mathrm{el}=e^{i\varphi_{1,2}}=\frac{i\omega-\Gamma_0/2}{i\omega+\Gamma_0/2},
\end{align}   
controlled by the energy of the photons in close analogy to the single polariton phase shift \eqref{eq:IndividualPhase}.

\begin{figure*}
  \centering
  \includegraphics[width=0.99\textwidth]{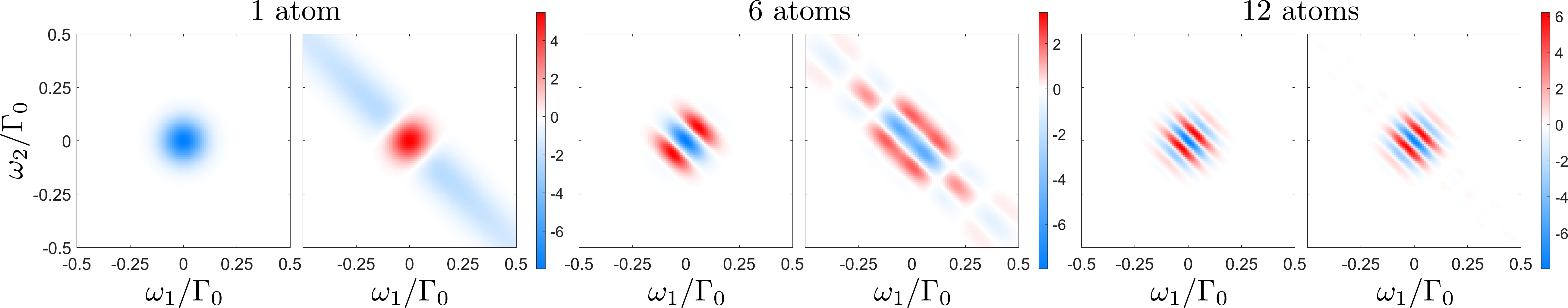}
  \caption{Real part of the two dimensional output state of an initial two-dimensional Gaussian state with width $\sigma=0.05\Gamma_0$ for both $\omega_1$ and $\omega_2$. For different emitter numbers (1,6,12) we compare our analytical approximation (left) with exact scattering matrix results (right).  }
\label{fig:numerics12}
\end{figure*}

In addition to the scattering phase, photons travelling to the right (left) will also acquire phases $\varphi_1$ ($\varphi_2$) due to the combination of propagation (6) and entering and leaving the medium for which  the mapping between photons and polaritons is given in Refs.  \cite{schrinski2021polariton,supp}. These phases are the same for all components of the state and thus only lead to an overall phase which can be ignored. The inelastic component acquire a more complex phase, but this will not influence the results below.

For the implementation of a gate we wish to suppress the inelastic scattering. For this reason we consider wave packets centered around the same detuning $\omega$, depending on the desired phase shift. This  i) suppresses the amplitude of inelastic scattering \eqref{eq:TransInelastic} and ii) ensures that inelastically scattered photons are far away in energy, c.f.  Eqs.~\eqref{eq:EnergiesInelasticScattering}.
The latter makes it easy to filter out these inelastic photons allowing us to absorb the inelastic scattering into a slightly decreased overall success rate of the gate. We note that alternatively one could try to maximize the inelastic scattering component by choosing $\omega_1=-\omega_2=\Gamma_0/2$ which produces  $t_\mathrm{el}=0$; the result is an energy swap between the two chiral modes, $\omega'_1=\omega_2$ and $\omega'_2=\omega_1$. Potentially this dynamics could also be used to implement quantum protocols.

\emph{Numerical verification--}The analytical results presented above describe the dynamics of an infinite chain. To investigate the performance for a finite chain we compare our results to exact scattering matrices derived from input-output theory \cite{caneva2015quantum}. In Fig.~\ref{fig:numerics12} we show the result of this for two Gaussian input states with widths $\sigma=0.05\Gamma_0$. 
Naturally, our approximation is worst for setups with very few emitters. Especially the high probability for inelastic scattering, as prominently indicated  by the ``jets" emerging on the sides of the initial state in Fig.~\ref{fig:numerics12} a) limits the performance of the gate with few emitters. These jets are completely gone for 12 emitters and furthermore the output states coincide quantitatively with the analytical results. In the latter case, the energy regions (sparsely) populated by inelastic scattering lie outside of the plotted area and are thus easily filtered out by frequency filters. 
The oscillations arise from the  phase \eqref{eq:IndividualPhase} accumulated by each photon.

\emph{Fidelity--}
To  characterize the performance of the  gate operation we exploit the Choi-Jamiołkowski isomorphism \cite{jamiolkowski1972linear,choi1975completely} and consider the input states to be  maximally entangled states 
\begin{align}
|\Phi\rangle\sim (|00\rangle_{12}+|11\rangle_{12})\otimes(|00\rangle_{34}+|11\rangle_{34}).
\end{align}    
Here, the gate only affects the qubits 1 and 3 while 2 and 4 denote auxiliary systems.
The Choi-Jamilkowski fidelity is then the overlap between the actual output state and the ideal gate operation, $\mathcal{F}=|\langle \Phi_\mathrm{ideal}|\Phi_\mathrm{out}\rangle|^2$. The fidelity will depend on which wave function we chose for the ideal state. Here we consider the mode  functions for the computational basis states $|0\rangle$ and $|1\rangle$ to be the dispersed states of the photons traversing the array individually, i.e.\, multiplying the input 2d wave function with $e^{iN(k(\omega_1)+k(\omega_2)-2k_0)d}$. In principle higher fidelity could be obtained by carefully optimizing this reference state. 

With this choice the fidelity  for perfect chirality is given by
\begin{align}
\mathcal{F}
=\frac{1}{4(3+t_\mathrm{norm})}|3+e^{-i\alpha}t_\mathrm{av}|^2
\end{align}
 for a desired phase $\alpha$, obtained by adjusting  $\varphi_{1,2}$ in Eq.~\eqref{eq:ScatteringPhaseshift}, and
\begin{align}
t_\mathrm{av}=&\int \mathrm{d}\omega_1\mathrm{d}\omega_2
|\psi_1(\omega_1)|^2|\psi_2(\omega_2)|^2t_\mathrm{el}(\omega_1,\omega_2)\nonumber\\
t_\mathrm{norm}=&\int \mathrm{d}\omega_1\mathrm{d}\omega_2
|\psi_1(\omega_1)|^2|\psi_2(\omega_2)|^2|t_\mathrm{el}(\omega_1,\omega_2)|^2.
\end{align}
Here, the  normalization in the denominator corrects for the success probability of the gate, assuming large detuning of the inelastic scattering so that it can be filtered out and  herald any failing of the gate.

\begin{figure}
  \centering
  \includegraphics[width=0.46\textwidth]{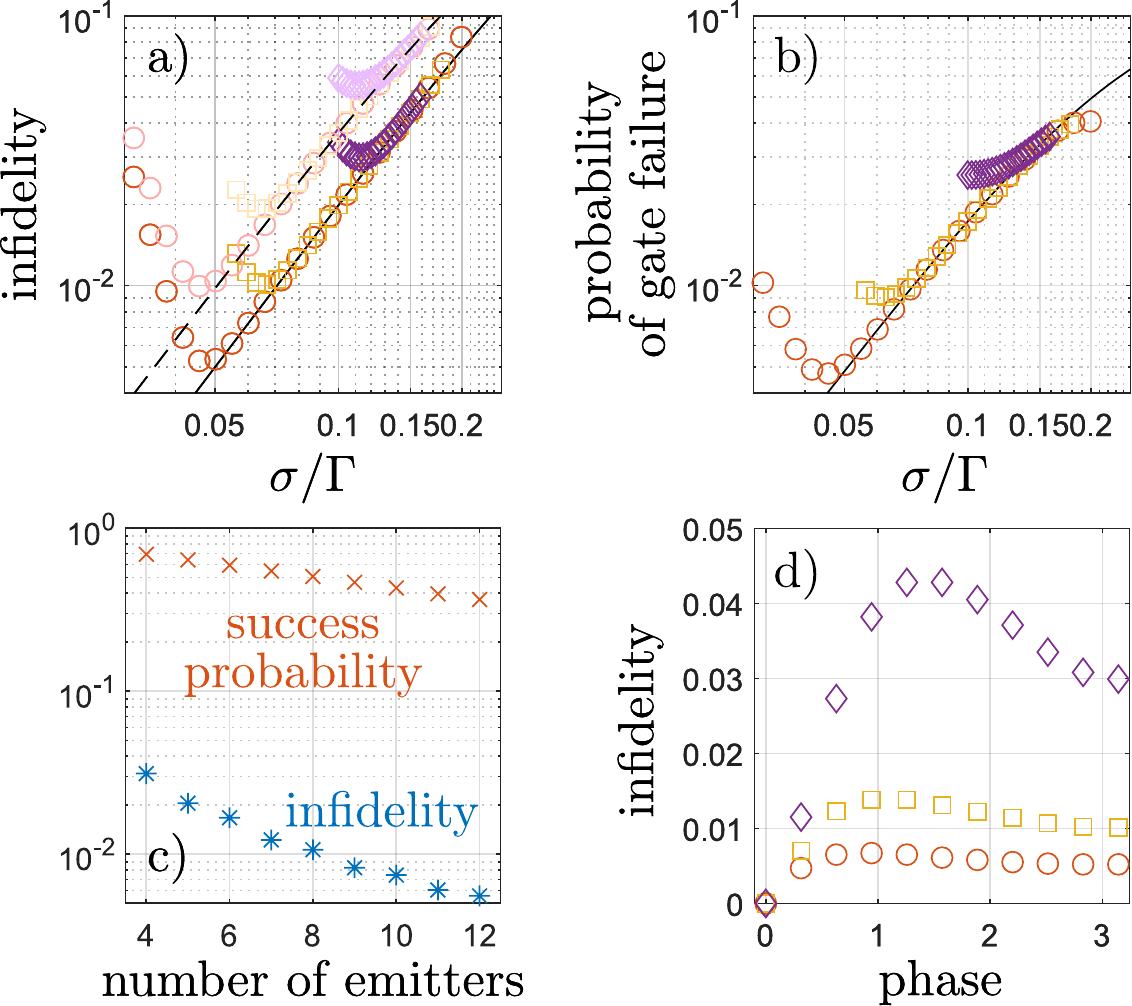}
  \caption{a) Infidelity $1-\mathcal{F}$ and b) probability of (heralded) gate failure as a function of the width $\sigma$ of an incident Gaussian pulse for a perfectly chiral setup. The solid line is for in infinite chain while the dark symbols represent  numerical results for 4 (diamonds), 8 (squares), and 12 (circles) emitters. The dashed line and lighter symbols in a) result from not filtering out the inelastic scattering. c) Imperfect coupling: Results assuming  that  photons are emitted backward or  to the side at identical rates rate $\Gamma_B=\Gamma_S=0.01\Gamma_0$. The success probability decays exponentially with the number of emitters while the fidelity still increases. We chose $k_0d=\pi$ to avoid Bragg scattering (see main text). d) Gate operation for arbitrary phase $\varphi$ by going  off-resonant~\eqref{eq:ScatteringPhaseshift}. The plot shows the lowest achievable infidelity for different numbers of 4, 8 and 12 emitters  using the same symbols as in a).
  }
\label{fig:FidelityProbability}
\end{figure}

\emph{Z-Phase gate--}As discussed above,  the scattering phase approaches minus unity if both photons are on resonance leading to perfect gate operation. This is illustrated in   Fig.~\ref{fig:FidelityProbability} a)$\&$b) where we show Fidelity and success probability  for an initial Gaussian distribution on resonance and varying width. As shown with lines  in the figure the analytical results for infinite chains rapidly approach the ideal limit as the incident states become narrow in frequency. For finite chains, we have excellent agreement as long as the pulses are sufficiently short in time (wide in frequency). As the pulses become wider in time, however, they can no longer fit inside the medium for a finite number of emitters. This leads to an optimal width of the incident pulses and a higher fidelity for a larger number of emitters. As a specific example, for 12 emitters and an initial width of $\sigma=0.05\Gamma_0$, a pulse width comparable to one recently used to for a related quantum dot experiment with a single emitter
\cite{chan2022chip},
 we have a near perfect fidelity $\mathcal{F}=99.48\%$ with only a minor reduction in success probability of the whole gate $\mathcal{R}=(3+t_\mathrm{norm})/4=99.51\%$.

If inelastic scattering events are not filtered out we obtain a success probability of unity but the infidelity is higher. As shown by the dashed line in Fig. ~\ref{fig:FidelityProbability} a),  $1-\mathcal{F}$ is doubled in the limit of many emitters. The breakdown for finite emitters occur at the same pulse width. Overall this thus leads to a factor of two increase in infidelity. 

The performance of the gate  relies on the chirality  of the setup. To investigate this we assume that photons are  emitted to the side or opposite to the intended direction in the waveguide at rates $\Gamma_S$ and $\Gamma_B$ respectively. The forward coupling is then charaterized by a coupling efficiency $\beta=(\Gamma_0+\Gamma_B)/\Gamma_\mathrm{tot}$ and directionality $D=(\Gamma_0-\Gamma_B)/(\Gamma_0+\Gamma_B)$, where $\Gamma_\mathrm{tot}=\Gamma_0+\Gamma_S+\Gamma_B$.  For the perfectly chiral case $\beta=1,D=1$  the positions of the emitters are completely irrelevant, but when $\Gamma_B\neq 0$ interference between scattering events introduce a dependence  on the exact placement.  
To minimize the effect of backscattering we assume $k_0 d=n\pi$ with $n\in \mathbb{Z}$. (Note that although this is equal to the usual condition for Bragg scattering, the strong chiral coupling modify the Bragg condition to occur at  $k_0=\pi/2+n\pi$ for our setup on resonance.) In  Fig.~\ref{fig:FidelityProbability} c) we present numerically optimized results for varying number of emitters and  $\Gamma_S=\Gamma_B=0.01\Gamma_\mathrm{tot}$ corresponding to $\beta=0.99$ and $D=0.98$. As seen in the figure non-perfect chirality reduces the success probability but the fidelity is almost unaffected. In practise this means that one will have to use a limited number of emitters for imperfect coupling, but it does not prevent the application of the gate.  

To investigate the effect of other imperfections we consider  a proof of concept realization with $N=4$ emitters  and parameters corresponding to  state of the art experiments with directionality and coupling of $D=0.98$ \cite{le2015nanophotonic} and $\beta=0.98$ \cite{arcari2014near}, respectively. In theory even higher values are possible  \cite{hauff2022chiral}. The full details are presented in Ref. \cite{supp}. As an example we incorporate an average fluctuation of $\Gamma_\mathrm{tot}$ with width $\sigma_{\Gamma_\mathrm{tot},\mathrm{dB}}=1.2\,$dB and of the resonance energy at each emitter with width $\sigma_\Delta=0.2\,\Gamma_\mathrm{tot}$. We also implement random distances $d$ between emitters and show that any mismatch in the pulse timing can be neglected for typical delay lines \cite{supp}. We average all fluctuating parameters over $10^3$ realizations resulting in  $\mathcal{F}>0.92$ and $\mathcal{R}>0.51$. For comparison, in  linear optics, CPhase gates are fundamentally limited by $\mathcal{R}\leq 1/9$ \cite{ralph2002linear}.

\emph{P-Phase gate--} The proposed setup is ideally suited  to induce a complete phase change by working  on resonance, which suppresses  inelastic scattering and minimizes the influence of finite numbers of emitters. 
According to  Eqs.~\eqref{eq:TransInelastic}$\&$\eqref{eq:ScatteringPhaseshift}  we can achieve a phase gate with an arbitrary  phase $\alpha$ by going off resonance with
\begin{align}
\omega_1=\omega_2=\frac{i}{2}\frac{1+e^{i\alpha}}{1-e^{i\alpha}},
\end{align}
while simultaneously suppressing inelastic scattering. According to the analytical approximation, assuming infinitely many emitters, this indeed leads to a perfect gate operation for any values of $\alpha$.

In reality, moving away from resonance jeopardizes the approximation of having many emitters since off-resonant photons interact less strongly   with each emitter. This leads to a higher group velocity, which in turn  means that we have to chose pulses more narrow in time to   localise the pulses inside the medium. This shifts the optimum pulse length to larger widths $\sigma$ in frequency for a finite number of emitters.
In Fig.~\ref{fig:FidelityProbability} d) we show the fidelity for different numbers of emitters and gate phases maximized over the energy width of the input states. As seen in the figure the infidelity increases as we go away from $\alpha=\pi$ until it again decreases as we approach $\alpha\rightarrow 0$, where the gates does nothing. 
For $N>10$ the infidelity is, however, on the permille scale for all values of $\alpha$.

\emph{Conclusion--}
We have presented a simple, completely passive approach to implement a gate between two photons. While the gate has ideal performance in the limit of many emitters it shows excellent behaviour even for a limited number of emitters.
The rapidly evolving field of chiral quantum optics promises proof of principle experiments with only a couple of emitters, which would already suffice to reach fidelities above 90$\%$. The developed photon gate could immediately be applied to facilitate photonic based quantum information processing, e.g.\, it allows the implementation of  efficient Bell-state measurements for photons  thereby enhancing the communication rate of ensemble based quantum repeaters \cite{sangouard2011quantum}.

\emph{Acknowledgements--}  
We are grateful to Bastian Bakkensen, Yuxiang Zhang, Peter Lodahl, G{\"o}ran Johansson, Per Delsing, Joshua Combes, Stefano Paesani, Alexey Tiranov, and Nils Valentin Hauff for useful discussions. 
B.\,S.\,is supported by Deutsche Forschungsgemeinschaft (DFG, German Research Foundation), Grant No. 449674892. We acknowledge the support of Danmarks Grundforskningsfond (DNRF 139, Hy-Q Center for Hybrid
Quantum Networks).

%\bibliography{Bibliothek}  


  \vspace{30pt}
 \appendix
\onecolumngrid
\begin{center}
\large{\textbf{Passive quantum phase gate for photons based on three level emitters: Supplemental material}}
\end{center}
 \vspace{10pt}
 
In the main text we give analytical results in the limit of (semi-)infinite arrays of emitters as an approximation for many emitters. We will provide the calculational details here and also discuss additional imperfections to the setup with minor implications. 

\section{Coupling of photons in and out of the atom array}

We divide the scattering process in three distinct parts \cite{schrinski2021polariton}: the individual (linear) in-coupling of photons to polaritons in a semi-infinite array, the subsequent (non-linear) scattering of two polaritons inside an infinitely extended medium, and finally the (linear) out-coupling  of the polaritons to photons at the boundaries of a semi-infinite array. Below we give the full details of the derivation of the in- and out-coupling as well as the scattering dynamics. Furthermore we present an investigation of the robustness of the gate to fluctuations in the parameters of the setup.

\subsection{Linear input and output}

We describe the coupling of the photons in and out of the medium in the linear regime where we consider only a single excitation at a time. Then the input-output formalism, exemplarily for a photon entering from the left chirally coupled to a semi-infinite chain, yields \cite{caneva2015quantum,schrinski2021polariton}
\begin{align}
\partial_t \sigma^\mu_{ge}(t)=-i\sqrt{\Gamma_0}\mathcal{E}_\mathrm{in}(z_1,t)e^{ik_0z_\mu}-\Gamma_0\sum_{\mu>\nu}e^{ik_0(z_\mu-z_\nu)}\sigma^\nu_{ge}(t)-\frac{\Gamma_0}{2}\sigma^\mu_{ge}(t)
\end{align}
with $z_\mu\in[0,\infty)$. We change to momentum space, $\sigma^k_{ge}(t)=\sqrt{d}\sum_\mu e^{-ikz_\mu}\sigma^\mu_{ge}(t)$, which yields
\begin{align}
\partial_t \sigma^k_{ge}(t)=&-i\sqrt{\Gamma_0d}\sum_{\mu=0}^\infty
\mathcal{E}_\mathrm{in}(z_1,t)e^{i(k_0-k)z_\mu}-\frac{\Gamma_0}{2}
\sigma^k_{ge}(t)-\sqrt{d}\sum_{\nu>\mu}\sigma^\mu_{ge}(t)\Gamma_0e^{ik_0|z_\mu-z_\nu|-ikz_\nu}\nonumber\\
=&\frac{-i\sqrt{\Gamma_0d}}{1-e^{i(k_0-k)d}}
\mathcal{E}_\mathrm{in}(z_1,t)-i\underbrace{\left(-\frac{\Gamma_0}{2}\frac{\cos[(k-k_0)/2]}{\sin[(k-k_0)/2]}\right)}_{=\omega_k}\sigma^k_{ge}(t).
\end{align} 
This equation can be solved in frequency space as \begin{align}
\sigma^k_{ge}(\omega)=\frac{-i\sqrt{\Gamma_0d}}{1-e^{i(k_0-k)d}}\frac{\mathcal{E}_\mathrm{in}(z_1,\omega)}{i(\omega_k-\omega)}
\end{align} 
and after back transformation we achieve via the Residue theorem
\begin{align}
\sigma^k_{ge}(t)=\frac{-i\sqrt{\Gamma_0d}}{1-e^{i(k_0-k)d}}\frac{1}{2\pi}\int_{-\infty}^\infty \mathrm{d}\omega\frac{\mathcal{E}_\mathrm{in}(z_1,\omega)}{i(\omega_k-\omega)}e^{-i\omega t}   
=\frac{-i\sqrt{\Gamma_0d}}{1-e^{i(k_0-k)d}} \mathcal{E}_\mathrm{in}(z_1,\omega_k)e^{-i\omega_k t},
\end{align}
representing a  1:1 mapping of input frequencies to polariton momentum states.

The output can be calculated with \cite{caneva2015quantum,schrinski2021polariton}
\begin{align}
\mathcal{E}_\mathrm{out}(z_N,t)=&-i\sqrt{\Gamma_0}\sum_\mu \sigma^\mu_{ge}(t)e^{ik_0|z_N-z_\mu|}
=\frac{1}{2\pi i}\sqrt{\Gamma_0d}\int_{-\pi/d}^{\pi/d} \mathrm{d}k\, \sigma^k_{ge}(t)\sum_{\mu=-\infty}^N e^{i(k-k_0)z_\mu}\nonumber\\
=&\frac{\sqrt{\Gamma_Rd}}{2\pi i}\int_{-\pi/d}^{\pi/d} \mathrm{d}k\,\frac{e^{i L (k-k_0)}}{1-e^{i(k_0-k)d}} \sigma^k_{ge}(t),
\end{align}
where we have used the inverse Z-transform  $\sigma^\mu_{ge}=\sqrt{d}\int_{-\pi/d}^{\pi/d} \mathrm{d}k\, \sigma^k_{ge}/2\pi$. 
We can once again do a transformation into frequency space
\begin{align}
\sigma^k_{ge}(\omega)=\lim_{\epsilon\to 0}\int_0^\infty \mathrm{d}t'\,e^{i\omega t'-\epsilon t'}\sigma^k_{ge}(t'),
\end{align}
to get
\begin{align}
\mathcal{E}_\mathrm{out}(z_N,\omega)
=&\lim_{\epsilon\to 0^+}\frac{\sqrt{\Gamma_0d}}{2\pi i}\int_{-\pi/2}^{\pi/d} \mathrm{d}k\frac{\sigma^k_{ge}(t=0)}{i(\omega_k-\omega)+\epsilon}\frac{e^{i L (k-k_0)}}{1-e^{i(k_0-k)d}}\nonumber\\
=&\lim_{\epsilon\to 0^+}\frac{\sqrt{\Gamma_0d}}{2\pi i}\int_{-\infty}^\infty \mathrm{d}\omega_k\frac{\sigma^-_{k(\omega_k)}(t=0)}{i(\omega_k-\omega)+\epsilon}\frac{e^{i L (k(\omega_k)-k_0)}}{1-e^{i(k_0-k(\omega_k))d}}\frac{1}{|v_{k(\omega_k)}|}\nonumber\\
=&-i\sqrt{\Gamma_0d}\frac{e^{i L (k(\omega)-k_0)}}{1-e^{i(k_0-k(\omega))d}}\frac{1}{|v_{k(\omega)}|}\sigma^{k(\omega)}_{ge}(t=0),
\end{align} 
where the group velocity $\partial_k \omega_k=v_k=\Gamma_0d/4\sin^2[(k-k_0)d/2]$ results from the Jacobian determinant, and the third line is achieved via the residue theorem. 

In the linear case the  in- and out-coupling transforms the electric field according to 
\begin{align}
\mathcal{E}_\mathrm{out}(z_N,\omega)
=e^{iN(k(\omega)-k_0)d}\mathcal{E}_\mathrm{in}(z_N,\omega).
\end{align}
The phase factor is simply the delay due to the absorption and re-emission by the emitters. On resonance ($k=k_0+\pi$) we obtain the alternating phase shifts $(-1)^N$.

\section{Two polariton scattering}

If we assume an array of many emitters we can approximate the non-linear scattering of two polaritons to occur deep inside the medium without being affected by the boundaries, i.e., we can assume an array of infinite extension.
The first step to solve the two polariton scattering is then to find the degenerate relative momenta of the two-photon dispersion relation which read
\begin{align}\label{eq:degenerateMomentum}
q'=-i\log\left[\frac{e^{ik_0d}(e^{ik_0d}+e^{i(k_0+2K)d}-2e^{i(K+q)d)}}{2e^{ik_0d}-e^{iqd}-e^{2i(K+q)d}}\right].
\end{align}
Plugging this back into the dispersion relation \eqref{eq:DispersionRelation} as $w'_1=w(K+q')$ (and analoguously $w'_2=w(-K+q')$) gives the inelastically scattered outgoing energies \eqref{eq:EnergiesInelasticScattering}. 

Next we want to evaluate the amplitudes $t_\mathrm{el}$ and $t_\mathrm{in}$ of elastic and inelastic scattering, respectively. To do this, we follow the procedure of Ref. \cite{Bakkensen2021photonic} and apply the effective Hamiltonian \eqref{eq:effHam} on the state \eqref{eq:effHamEigenstate} resulting in 
\begin{align}
\mathsf{H}|\psi_{q,K}\rangle
=&\omega_{q,K}|\psi_{q,K}\rangle+(a(q-K-k_0)+a(-q+K+k_0)t_\mathrm{el}+a(-q'+K+k_0)t_\mathrm{in})|\psi_{2k_0+2K,K}\rangle\nonumber\\
&+(a(q+K-k_0)+a(-q-K+k_0)t_\mathrm{el}+a(-q'-K+k_0)t_\mathrm{in})|\psi_{2k_0-2K,K}\rangle,
\end{align}
with $a(\kappa)=1/(e^{i\kappa d}-1)$. The two states with relative momenta $2k_0\pm 2K$ need to have vanishing prefactors in order to ensure that  $|\psi_{q,K}\rangle$ is a proper eigenstate. This leads to two independent equations to obtain the two scattering amplitudes. Plugging in the degenerate momentum \eqref{eq:degenerateMomentum} then leads to
\begin{align}
t_\mathrm{el}=&\frac{e^{2i(k_0-q)d}-2e^{i(k_0-q)d}\cos[Kd]+\cos[2Kd]}{2\cos[Kd]\cos[(k_0-q)d]-2}\\
t_\mathrm{in}=&\frac{2(\cos[Kd]-\cos[(k_0-q)d])(\cos[(k_0+K)d]+i\sin[(k_0+K)d])\sin^2 [Kd]}{(e^{ik_0d}+e^{i(k_0+2K)d}-2e^{i(K+q)d})(\cos[Kd]\cos[(k_0-q)d]-1)}.
\end{align}
Inverting the dispersion relation \eqref{eq:DispersionRelation} and substituting $K=(k_1+k_2)/2$ and $q=(k_1-k_2)/2$ reveals the scattering amplitudes as functions of the input frequencies as given in \eqref{eq:TransElastic} and \eqref{eq:TransInelastic} of the main text.

\begin{figure}[b]
  \centering
  \includegraphics[width=0.99\textwidth]{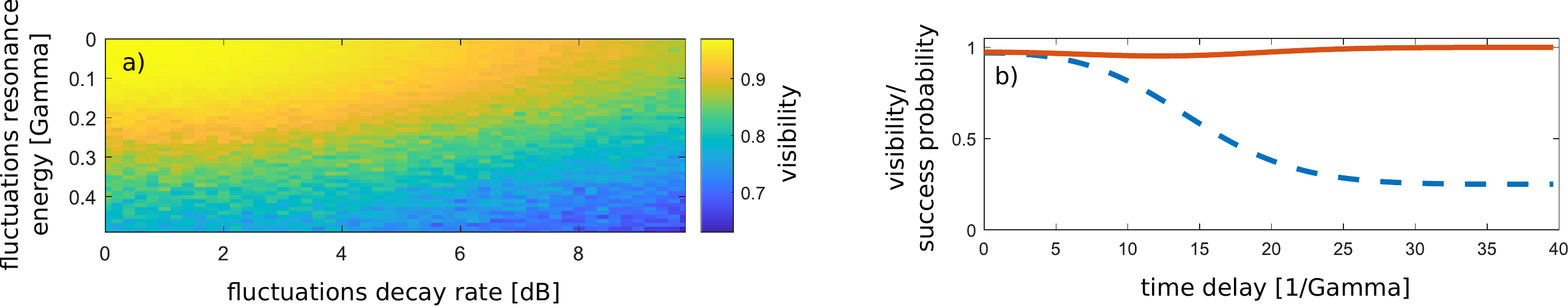}
  \caption{a) Dependence of the fidelity  on deviations  in decay rates and resonance energies among the various emitters in the array. The axes represent the standard deviation of fluctuations, the first  in units of the decay rate $\Gamma$, the latter  in Decibel. b) Dependence of  the fidelity (dashed curve) and success probability (full curve) of the gate operation on potential mismatch in arrival time between the photon pulses. In the limit of completely unaligned pulses the fidelity assumes its minimum value $0.25$ while the success probability becomes unity since no inelastic scattering takes place. In both plots a) and b) we have $N=4$ and no other imperfections present apart from the specific ones considered.
  }
\label{fig:Supp1}
\end{figure}

\section{Imperfections}

In the main text we discuss the most important sources of imperfections affecting the quality of the gate, namely the  directionality of the chiral coupling and fraction of coupling into the waveguide as well as the placement of the emitters. More forgiving are fluctuations in the decay rate and resonance energy. For the former we incorporate relative fluctuations of the decay rate of the $\mu$th emitter in the form of $\Gamma_\mu=\Gamma_0 10^{\mathcal{N}(0,\sigma_{\Gamma_\mathrm{tot,dB}})/10}$ with a normally distributed \emph{relative} fluctuation on the Decibel scale $\mathcal{N}(0,\sigma_{\Gamma_\mathrm{tot,dB}})$ centered around zero. For the fluctuations in resonance energy of the $\mu$th emitter we introduce absolute fluctuations of the form $\Delta_\mu=\Gamma_0\mathcal{N}(0,\sigma_{\Delta})$. The impact of both is depicted in Fig.~\ref{fig:Supp1} a). From the figure it is seen that fluctuations in the decay rates of up to $\sim 5$ dB and resonance energies up to $\sigma_{\Delta}\sim 0.2 \Gamma$ have little influence on the resulting fidelity. 

A mismatch between arrival times of the photon pulses in form of a delay time, depicted in  Fig.~\ref{fig:Supp1} b), turns out to also be of limited consequence: Since the optimal pulse duration for $N=4$ is  around $\sim 9/\Gamma$,  the fidelity is only affected by syncronization mismatch around several times the life time $1/\Gamma$ of the quantum dot or larger. Taking quantum dots as an example, typical emitter life times can be on the order of 400 ps \cite{chan2022chip}. The time to propagate through the ensemble is set by the Wigner delay $4N/\Gamma$ \cite{mahmoodian2020dynamics}. This means that the typical delay between  pulses can be $\sim 10 N/\Gamma$. For $N=4$ this corresponds to pulses of duration 3.6 ns with 16 ns delay between the pulses equivalent to delay lines of length $40 c/\Gamma\sim 4$ m,  
which should be adjusted with a precision of around $\sim 0.5$ m. For comparison a recent related scattering experiment with a single quantum dot with a lifetime of 0.4 ns employed pulses with a 2 ns duration and 12 ns delay \cite{chan2022chip}.

\end{document}